\documentstyle{mn}
\begin{document}

\def\Ej{{\rm E}_{\rm J}}
\def\OrbFam#1#2#3{$#1\!:\!#2\!:\!#3$}
\def\empty#1{\vbox to#1{\rule{0pt}{#1}}}

\def\plotfiddle#1#2#3#4#5#6#7{\centering \leavevmode
\vbox to#2{\rule{0pt}{#2}}
\includegraphics{#1}}

\title[Gas-driven evolution in barred galaxies]
      {GAS-DRIVEN EVOLUTION OF STELLAR ORBITS\\
       IN BARRED GALAXIES}

\author[I. Berentzen, C.H. Heller, I. Shlosman and K.J. Fricke]
       {I. Berentzen,$^1$ C.H. Heller,$^1$ I. Shlosman$^2$
         and K.J. Fricke$^1$\\
        $^1$Universit\"ats Sternwarte, Geismarlandstra\ss e 11, 
            D-37083 G\"ottingen, Germany \\
        $^2$Department of Physics and Astronomy, University of Kentucky,
            Lexington, KY 40506-0055, USA}

\maketitle

\begin{abstract}

We carry out a detailed orbit analysis of gravitational potentials 
selected at different times from an {\it evolving} self-consistent 
model galaxy consisting of a two-component disk (stars$+$gas) and a live
halo. The results are compared with a pure stellar model, subject to 
nearly identical initial conditions, which are chosen as to make the models
develop a large scale stellar bar. 
The bars are also subject to hose-pipe (buckling) instability which
modifies the vertical structure of the disk. The diverging morphological
evolution of both models is explained in terms of gas radial inflow, the
resulting change in the gravitational potential at smaller radii, and the
subsequent modification of the main families of orbits, both in and out
of the disk plane.

We find that dynamical instabilities become milder in the presence of
the gas component, and that the stability of planar and 3D stellar orbits is 
strongly affected by the related changes in the potential --- both are
destabilized with the gas accumulation at the center. This is reflected
in the overall lower amplitude of the bar mode and in the substantial 
weakening of the bar, which appears to be a gradual 
process. The vertical buckling of the bar is much less pronounced and
the characteristic peanut shape of the galactic bulge almost
disappears when there is a substantial gas inflow towards the center.
Milder instability results in a smaller bulge whose basic parameters
are in agreement with observations. 
We also find that the overall evolution in the model with a gas component 
is accelerated due
to the larger central mass concentration and resulting decrease in the
characteristic dynamical time.

\end{abstract}


\section{INTRODUCTION}

Our understanding of dynamics in disk galaxies is hampered by the
prevalence of non-axisymmetric features there, most notably stellar bars. 
Formation and evolution of barred disks appears to be a formidable problem
whose solution continues to elude us (e.g., Lynden-Bell 1996).
Galactic structure and stability are traditionally studied in terms
of dominant stellar orbits
(Sellwood \& Wilkinson 1993, and refs. therein). Recently, it became clear
that major families of stellar orbits are profoundly influenced by 
dynamical and secular effects in the galactic disks, the latter on 
timescales shorter than the Hubble time. This evolution is driven
in part by dynamical instabilities in the stellar component, such as the
bar instability in the disk plane (e.g., Ostriker \& Peebles 1973)
and the bending instability out of the equatorial plane (Toomre 1966;
Combes et al. 1990; Pfenniger \& Friedli 1991; Raha et al. 1991; etc.).
The importance of the three-dimensional nature of
the stellar bars for the formation of exponential disks and galactic bulges
is now recognized (e.g., Pfenniger 1984; Pfenniger \& Friedli 1991;
and others).

An additional driver of orbital evolution is the disk gaseous component
whose mass, although small compared to the overall galactic mass, can
nevertheless exert dynamical effects on stars and substantially modify
galactic morphology.
First, the cold interstellar medium (ISM) is clumpy, with the mass spectrum 
weighted towards the higher masses, and with the high-mass cutoff around 
$10^7\ M_\odot$ \cite{san85}. Such inhomogeneities efficiently and 
randomly scatter stars residing on periodic and semi-periodic orbits, 
resulting in overall disk heating, and may even impede the bar instability,
reducing the overall phase space available to these orbits \cite{shl93}. 

Second, the ability of the gas to dissipate its rotational energy is expected
to lead to a radial redistribution of the ISM in the disk. Within the 
corotation, the gas flow is directed inwards as the gas loses its angular
momentum support dynamically, through gravitational torques and large
scale shocks in the bar, or secularly, in the spiral arms. Gas inflow to the
center can result in a substantial change in the galactic gravitational
potential at small radii and is likely to be accompanied by star formation
and formation/fueling of an active galactic nucleus \cite{shl89}.
The 2D and 3D orbit analysis in {\it static} or {\it partially static}
potentials has shown that growing 
central mass concentrations, like compact bulges, nuclear star clusters,
or supermassive black holes, all tend to destroy the main family of
periodic orbits aligned with and supporting the stellar bar
(Hasan \& Norman 1990; Hasan, Pfenniger \& Norman 1993;
Norman, Sellwood \& Hasan 1996). 
This leads to a subsequent weakening of the bar, and 
ultimately to its `quick' dissolution. The same trend but with clear
differences was found when a massive nuclear molecular ring is present in 
a starburst galaxy, preferentially destroying stable orbits between the
inner Lindblad resonance(s) (ILRs) and the corotation \cite{hel96}. 

In this work we take the next logical step by performing a 2D and 3D
orbit analysis at selected times of an
{\it evolving} and fully self-consistent galactic potential
generated by a disk embedded in a live halo. We provide a comparison 
between pure stellar and two-component stars$+$gas models using nearly
identical initial conditions, so that the different evolutionary paths can be
attributed to the presence of the dissipative gaseous component. The model
axisymmetric galactic disk is chosen as to experience both bar and bending
instabilities.
We aim at understanding to what degree the gaseous component affects these
instabilities and modifies the main families of orbits both in and out of the 
equatorial plane. By doing so, we follow in the footsteps of Sparke \&
Sellwood \shortcite{spa87} and Pfenniger \& Friedli \shortcite{pfe91},
whom analyzed the stellar orbits of self-consistent time-dependent
2D and 3D potentials, of pure stellar disks. The main modification
introduced in the present work is the addition of a dissipative
gravitating component in the disk and a responsive `live' halo.

Section~2 describes the numerical algorithm used to evolve the 
multicomponent galaxy model. The model initial conditions are given 
in section~3, and section~4 deals with the evolution. The orbit analyses 
for the pure stellar and stars$+$gas models are given in Section~5. 
The main results are discussed in Sections~6 and 7.

\section{METHOD}

The method consists of an N-body algorithm to evolve the collisionless
component, representing the stars and dark matter, combined with a 
smoothed particle hydrodynamics (SPH) algorithm to evolve the
dissipative component, representing the gas (e.g., review by Monaghan 1992).  

In SPH, as in grid-based codes, the continuous physical fields
(e.g. density and velocity) are approximated by a set of points and
smoothed quantities are obtained by averaging over finite volumes.
However, in SPH these interpolating points move with the mean fluid
velocity and the averages are computed using a kernel weighting
function.  

The algorithm, employs such features as, a spatially varying smoothing 
length, a hierarchy of time bins to approximate individual
particle timesteps, a viscosity ``switch'' to reduce the effects of
viscous shear, and the special purpose GRAPE-3Af hardware to compute
the gravitational forces and the neighbor interaction lists 
(Sugimoto et al. 1990; Steinmetz 1996).  
Further details and tests of this algorithm can be found in 
Heller \shortcite{hel95} and Heller \& Shlosman \shortcite{hel94}.

For the orbit analysis we use the algorithm described in Heller \&
Shlosman \shortcite{hel96}.
The potential is prepared by evaluating it on a rectangular
three-dimensional grid, from the particle model using the GRAPE hardware.
The grid spacing is adjusted with position in order to give an
appropriately smooth field without losing relevant features, such
as the bar, gas ring and inner disk.  The potential is then symmetrised
with eight-fold symmetry.

The potential and its derivative 
are evaluated using the B3VAL routine from the CMLIB package.  This
routine determines and evaluates a a piecewise polynomial function
represented as a tensor product of one-dimensional B-splines. 
We have checked the suitability of
this technique for our application in several ways, including comparing
the mass density, as evaluated from Poisson's equation, to the 
density determined directly from the particle model.  The computations
were carried out on local workstations and a CrayT3D in Berlin.  

\section{MODEL}

The initial density distribution for both the pure stellar model (A)
and the stellar+gas model (B) are derived from the
Fall \& Efstathiou \shortcite{fal80} disk-halo analytic model which
consists of an exponential disk and spherical halo.
We adopt units for mass, distance, and time of, respectively,
$M=10^{11}\,M_{\sun}$, $R=10$\,kpc, and 
$\tau = \tau_{\rm dyn} \equiv (R^3/GM)^{1/2} = 4.7\!\times\!10^7$\,yr.
The initial conditions are such that within one unit of distance (10\,kpc)
both the total mass and the ratio of disk-to-halo mass are unity. This makes
the initial period of rotation at 10\,kpc to be 
$t_{\rm rot} \equiv 2\pi\tau_{\rm dyn}$.
The radial and vertical exponential scale heights in the disk are 
2.85\,kpc and 0.2\,kpc, respectively, and the rotational velocity 
turnover radius is $r_{\rm m} = 7$\,kpc.  
The halo is initially populated with 30K collisionless
particles within a radius of 30\,kpc, while the disk has 100K particles 
within 25\,kpc.  A gravitational softening length of 0.16\,kpc is
used for all particles.

To check the degree to which the more massive halo particles may heat
the stellar disk, a model with the softening length reduced by a factor
of two was also constructed.  The evolution of the two models were 
essentially identical and we therefore conclude that two-body heating
of the disk by the halo is not significant.

For model~B, some 10K collisionless disk particles are replaced with 
collisional SPH particles with a vertical scale height of 0.15\,kpc, 
representing 8\% of the global mass within 10\,kpc.
Also a central object is added which absorbs all particles
within a radius of 40\,pc and its mass grows at their expense.  An 
isothermal equation of state with
a temperature of $10^4$\,K is used for the gas.

This distribution is not in an exact virial equilibrium and the halo
must be allowed to relax from its initially spherical shape.  Details
of this process along with the assigning of velocities and a
description of the resulting density and velocity profiles can be found
in Shlosman \& Noguchi \shortcite{shl93} and Heller \& Shlosman
\shortcite{hel94}.

\section{MORPHOLOGICAL EVOLUTION}

The model was constructed so as to be globally unstable to non-axisymmetric
perturbations and form
a large-scale bar in a few $t_{\rm rot}$.  The bar itself is subject to
bending instability which changes the vertical structure of the 
stellar disk on a dynamical timescale.  In this section we look at 
how the presence of gas changes the character of these instabilities,
along with the overall secular evolution of the model galaxy.
This evolution is shown in Figures~1 and 2, and examples of dominant
orbits appear in Fig.~3.
\begin{figure*}
\plotfiddle{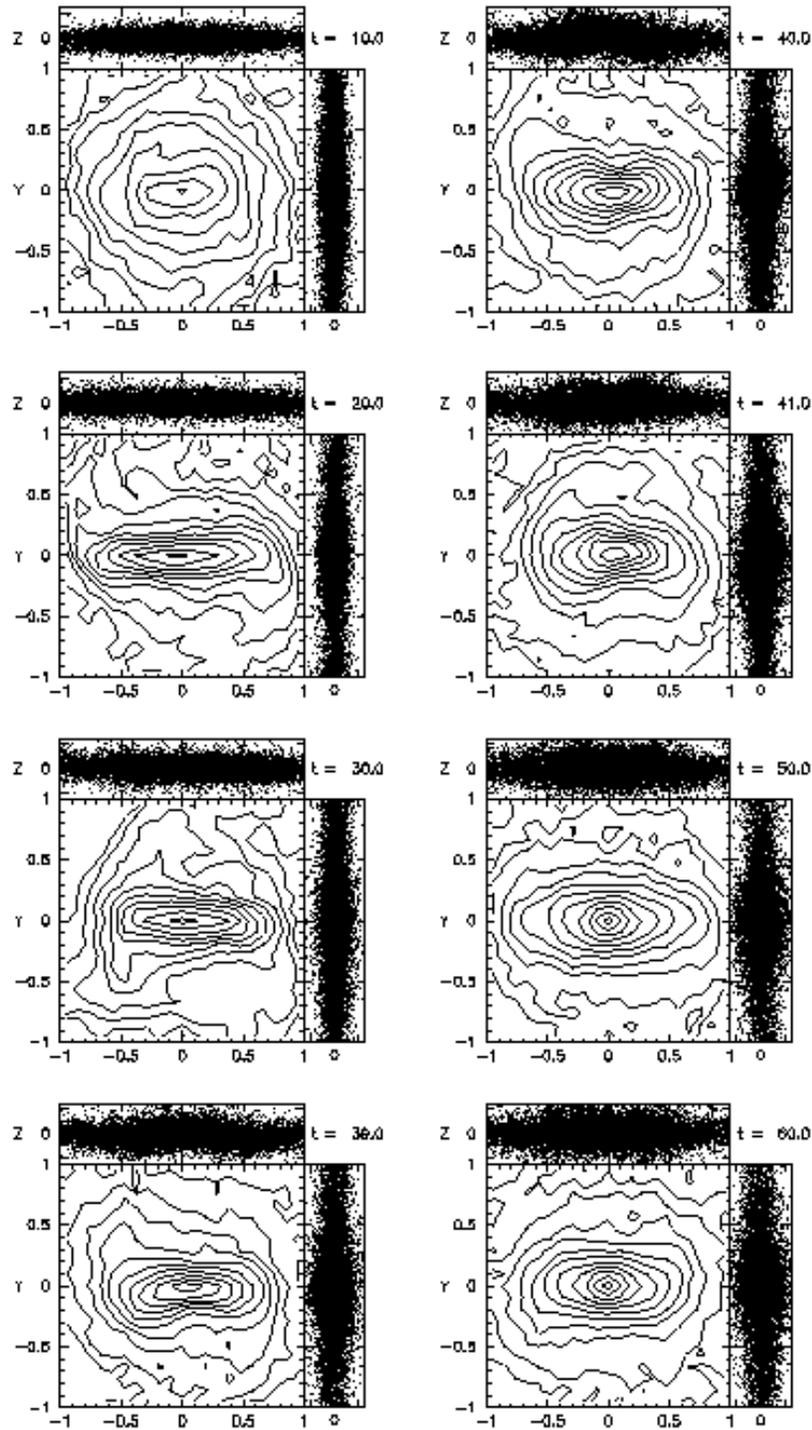}{21cm}{0}{90}{90}{-275}{-70}
\caption{Evolution of stellar disk in model~A rotating counter-clockwise.  
Shown are face-on
stellar contours and edge-on stellar particle distribution.  For
clarity, only 1/4 of disk stellar particles are plotted.  The disk
has been rotated in each frame so that the bar is along the x-axis.}
\end{figure*}
\begin{figure*}
\plotfiddle{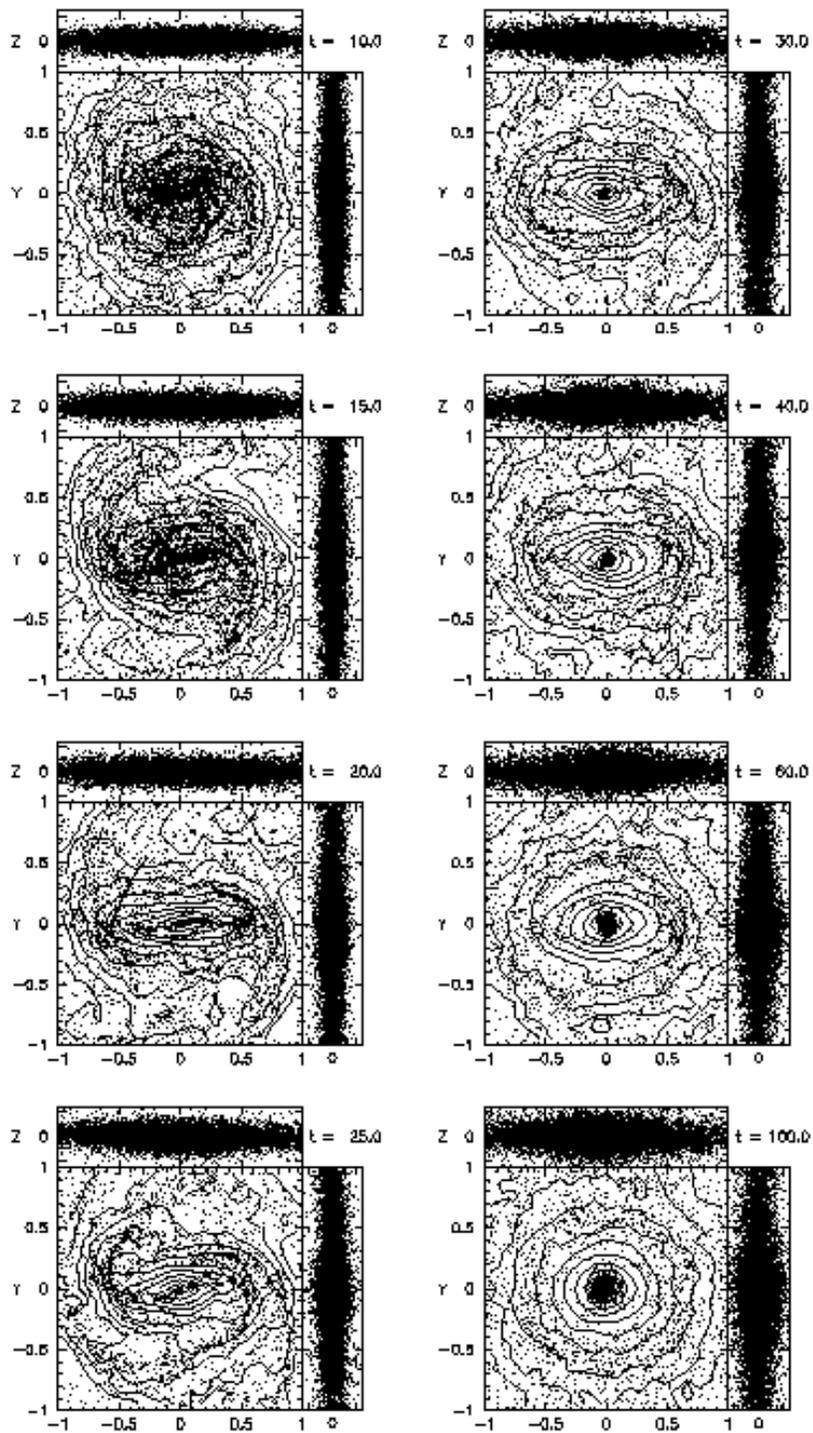}{21cm}{0}{90}{90}{-275}{-70}
\caption{Evolution of stellar+gas disk in model~B rotating counter-clockwise.  
Shown are face-on
stellar contours, face-on sph particle distribution, and edge-on stellar
particle distribution.  For clarity, only 1/4 of disk stellar particles
are plotted.  The disk has been rotated in each frame so that the bar
is along the x-axis.}
\end{figure*}
\begin{figure*}
\plotfiddle{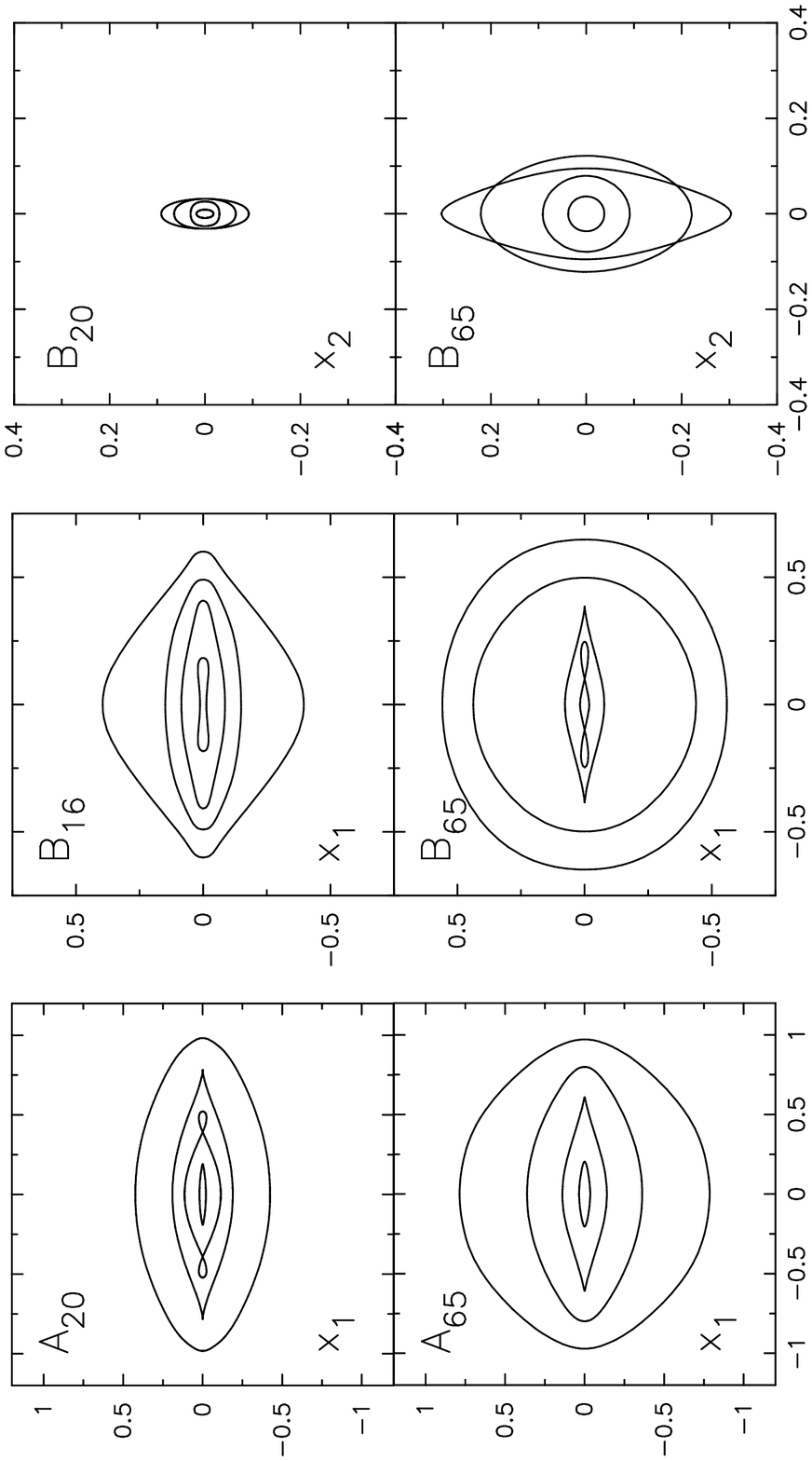}{7cm}{-90}{50}{50}{-180}{200}
\caption{Some examples of $x_1$ and $x_2$ planar orbits from models
A and B.}
\end{figure*}

\subsection{Model Without Gas}

For the purely collisionless model~A, the bar
forms and then reaches maximum strength at about $t=20$ or 
$3.2\,t_{\rm rot}$.
At this time the bar extends out to about 8\,kpc with axis ratios
of approximately (x:y:z) 1.0:0.3:0.13, as estimated by eye from
the density isophotes.  From the ends of the bar 
a trailing spiral emerges, which reaches a maximum radius of some 
14\,kpc at an azimuth 
perpendicular to the bar major axis.  These arms dissolve slowly and
have completely disappeared by $t=60$ or after some 8 bar rotations.
The pattern speed of the bar also decreases during this interval,
from a maximum of $\Omega_{\rm b} = 1.08\,{\rm rad}/\tau$ at a
time shortly before bar maximum strength ($t=20$), to 
$\Omega_{\rm b} = 0.82\,{\rm rad}/\tau$ at $t=65$. This decline
is most rapid during the time the bending instability is operating and
then slows to a steady rate of about
$\Delta\Omega_{\rm b}=-0.002\,{\rm rad}/\tau^2$, 
implying a slowdown timescale greater than a Hubble time.
No radial ILRs exist in this model at 
any time as the orbit analysis of section~5.1 shows, although linear analysis
claims (erroneously, due to the bar strength) a double ILR in the disk plane.
Both linear and orbit analyses confirm the existence of a single vertical ILR. 

Immediately following maximum strength, the bar begins to weaken
and at around $t=35$ a vertical buckling becomes visibly discernible
as the bar has lost its symmetry with respect to the equatorial $z=0$ plane.
The amplitude of this vertical asymmetry in the bar grows considerably, 
until at $t\sim45$ the bar takes on a boxy appearance. The peanut shape of disk
viewed along the bar's minor axis is clearly visible during this instability,
while the bar appears round when seen along its major axis. After about
$t=60$ the bar remains stable and any evolution it is undergoing is 
taking place on a much longer timescale.  While the bar length has not
changed appreciably during the evolution, the axis ratios have become
1.0:0.6:0.18, indicating both the weakening of the bar in the plane
and the vertical thickening.

Analysis of the vertical distribution of the stellar disk during
the evolution shows that the bending instability leaves behind  a ring 
of enhanced vertical scale height.  Before the instability the
stellar disk scale height is at a minimum at the disk center and
increases outward as the disk flares.  Following the instability
the scale height over the entire disk has increased, but particularly
so along a circular shaped ridge at a radius of approximately 3--4\,kpc.

We have examined the disk and bulge radial scale lengths of
the model by fitting to the stellar surface density the function
\begin{equation}
\Sigma(r) = \Sigma_{\rm d} \exp\left[-\frac{r}{r_{\rm d}}\right]
	  + \Sigma_{\rm b} \exp\left[-\left(\frac{r}{r_{\rm b}}
			   \right)^{1/n}\right].
\end{equation}
The shape parameter, $n$, for the bulge component corresponds to
a pure exponential profile for $n=1$ and a deVaucouleurs profile
for $n=4$.  Initially, at $t=0$,  the distribution is a pure
exponential with $r_{\rm d}=2.86$\,kpc, while at $t=65$ both
components are required for an adequate fit, as confirmed by
an F-test.  The parameter values at $t=65$ are $r_{\rm d}=2.18$\,kpc,
$r_{\rm b}=0.76$\,kpc, $n=0.66$, and $\Sigma_{\rm b}/\Sigma_{\rm d}=2.42$.
In the $r_{\rm b}/r_{\rm d}$ vs. $n$ plane these values
are on the outer edge of the observed distribution, albeit not
strictly incompatible given the large amount of observational
scatter \cite{and95}.  Fixing $n=1$ we find a bulge-to-disk ratio
$r_{\rm b}/r_{\rm d}=0.18$ which is about two-sigma above the mean
(Courteau et al. 1996; Courteau 1997).

We have also investigated the angular momentum loss of the inner
disk to the outer disk and the halo.  We find that the disk within
corotation loses, by $t=65$, about 40\% of its initial angular 
momentum, with 70\% of this going to the disk outside of corotation
and the remaining 30\% to the halo, with the outer halo gaining
about twice as much angular momentum as the inner halo.
The loss of angular momentum from the inner disk occurs fairly
secularly, except for a burst of angular momentum transfer to the
outer disk during the epoch of bar formation ($t=$14--20).

\subsection{Model With Gas}

In model B, which includes gas, the bar reaches maximum strength
at about the same time ($t=20$) as the purely collisionless model~A.
At this time the stellar distributions are nearly identical in the
two models, while the gas in model B
has formed a strong shock along the bar, offset in the leading
direction and curving inward near the center.  This gas morphology 
is indicative of the presence of an ILR \cite{ath92},
as is confirmed by our orbit analysis. 

Even at this early time there has been a substantial
inflow of gas, with some 40\% of the gas that was initially within
10\,kpc, now residing within a radius of 1\,kpc. This gas represents
some 23\% of the dynamical mass within the inner kiloparsec.
In fact about 80\% of this gas mass has gone to the inner few hundred
parsecs and resides in the central accreting object, which contains
at this time about 1.6\% of the total galactic mass within 10\,kpc.
Following this large burst of mass accretion by the central object,
it continues to grow linearly over time, reaching by the end of the
run, at $t=100$, $2.3\!\times\!10^9\,M_{\sun}$ or 2.2\% of the global
mass inside 10\,kpc.  This corresponds to a gas density increase by a
factor of $\sim6.5$ within the central kiloparsec.

As before, the stellar arms slowly dissolve until they are not
visibly discernible after about $t=60$.  However, large spiral
features in the gas persist throughout the run.  The pattern speed of the bar
decreases during this interval from a maximum of 
$\Omega_{\rm b} = 1.31\,{\rm rad}/\tau$ (at $t=16$) to a minimum of
$\Omega_{\rm b} = 1.13\,{\rm rad}/\tau$ at around $t=65$, after which it
remains constant or possibly is even very slowly increasing, in agreement with
previous work (Heller \& Shlosman 1993, unpublished).  

By $t=30$ the gas circular velocities have developed a sharp
discontinuity in the slope at about 1\,kpc from the center. 
The gas rotation increases up to this radius
and thereafter remains approximately flat throughout the bar region. 
The central 300\,pc are dominated by a
growing oval gas disk, whose major axis leads the stellar bar by
$\sim 80^\circ$. The gas,
in fact, is accumulating close to the inner ILR (as the orbit analysis of
section~5 confirms). Outside this disk is a noticeable deficiency of gas in
the bar, up to the radius of a forming oval
ring of gas which surrounds the stellar bar at about the position
of the Ultra-Harmonic Resonance (UHR).  Model~A has the rectangular outer
isophotes (Fig.~1) characteristic of a strong UHR, while in model~B they
are somewhat weaker (Fig.~2).  The inner gas disk
and the UHR gaseous ring remain throughout the run, connected by
thin trailing gas spiral shocks, offset from the stellar bar in 
the leading direction.  The inner gas disk continues to grow in size, being
fed through the shocks and reaching a radius of approximately 1.2\,kpc 
by $t=100$. We note that a similar gas morphology was obtained in the
numerical modeling of NGC~4321 by Knapen et al. \shortcite{kna95}.

Similar to Model~A, the bar begins to weaken after reaching maximum
strength,  but in this case it continues to do so, linearly with time
until the end of the run.  A vertical bending of the stellar disk
also occurs, but earlier, near the time of maximum bar strength.
The bending is not as dramatic as before, and in fact is difficult
to detect from a visual inspection of the stellar distribution.
The jump in vertical scale heights at the time of the instability 
is also neither quite so abrupt or substantial.  The gas acts 
to weaken the instability.  The ridge-like 
feature that was seen in the purely collisionless model following the 
instability is also present here but at a slightly smaller radius (2--3\,kpc).

The bulge/disk parameters, as given by a fit of the model to equation (1),
are at $t=65$:
$r_{\rm d}=2.87$\,kpc, $r_{\rm b}=0.28$\,kpc, $n=1.26$, 
$\Sigma_{\rm b}/\Sigma_{\rm d}=13.80$,
and at $t=100$:
$r_{\rm d}=3.05$\,kpc, $r_{\rm b}=0.20$\,kpc, $n=1.53$, 
$\Sigma_{\rm b}/\Sigma_{\rm d}=21.08$.  
Using these scale lengths and integrating out to $5r_{\rm d}$ we
find at $t=100$ a bulge-to-disk mass of ${\rm M_b/M_d}=0.3$.
These values fit well with the observed $r_{\rm b}/r_{\rm d}$ vs. $n$
relationship. Hence in relation to model~A, the presence of the gas has
moved the values in the direction towards better compliance with
the observed distribution.  Also, the values of $n$ are a factor of two
or more greater than that of model~A, consistent with an
evolution in the direction of earlier spiral-type \cite{cou97}.
Fixing $n=1$ we find $r_{\rm b}/r_{\rm d}=0.15$, which is lower
than that found for model~A, but still somewhat on the high end of
the observed distribution \cite{cou96}.

The angular momentum loss of the inner disk, by $t=65$ is some 10\% less
than that of model~A, though the percentages absorbed by the halo
and outer disk are roughly equivalent.
In spite of the large inflow of gas, the overall behavior of the
angular momentum redistribution during the evolution is quite similar to
that of the purely collisionless model.

\section{ORBITAL EVOLUTION}

In this section we examine the evolution of the orbits in the 
models. General information on orbits in barred galaxies can be found
in Binney \& Temaine \shortcite{bin87} and Sellwood \& Wilkinson
\shortcite{sel93}. We do so by locating the periodic orbits, 
i.e. orbits which
make a closed figure in a frame of reference that rotates with the bar,
in the frozen potential at a given time.  For simplicity, we restrict
ourselves to only the lowest order periodic orbits in the symmetrised
potential within the corotation radius.  We compare both the planar 
and 3D orbits, from each model at two different
times.  The first time was chosen to be when the bar is near maximum strength,
but before the onset of the bending instability, the second after the
instability, at a time when the evolution has reached a quasi-static state.

We start by searching for simple 1-periodic orbits in the plane 
($z=\dot{z}=0$), that is planar orbits which are bi-symmetric with
respect to the bar and close after one orbit around the center in the
rotating frame of reference.  The stability of these orbits is computed
and orbits which bifurcate in $z$ and $\dot{z}$ from vertically unstable 
regions are then located.  The results are displayed in terms of a 
characteristic diagram, where the orbits are plotted with respect
to their Jacobi Integral, $\Ej$, and either the $y$, $z$, or $\dot{z}$
intercept value with the $x=0$ plane.  The Jacobi Integral (or Energy) is
a conserved quantity along any given orbit in the rotating frame,
and can be thought of as an effective energy.  In the characteristic
diagrams the orbits form curves or families.  It is the study of
these families, their properties and how these change during the 
evolution of the models, that concerns us in this section.

A number of different notations to designate the orbital families
have been used in the literature.  Unfortunately all have some 
drawback, either in terms of being limited in the types of orbits
covered or in being unfamiliar to most readers.  We adopt here
a mixed approach.  For the planar orbits we will take the widely
used notation of Contopoulos \& Papayannopoulos \shortcite{con80},
while for the more common 3D orbits the Geneva system will be used.
For the few 3D orbits described here which do not have a Geneva designation
the notation of Sellwood \& Wilkinson \shortcite{sel93} is used.
The correspondences between the systems will also be noted.

\subsection{Model Without Gas}

\subsubsection{Periodic Orbits at $t=20$}

The characteristic diagram for the planar orbits of Model~A at $t=20$
is shown in Figure~4a. 
\begin{figure*}
\plotfiddle{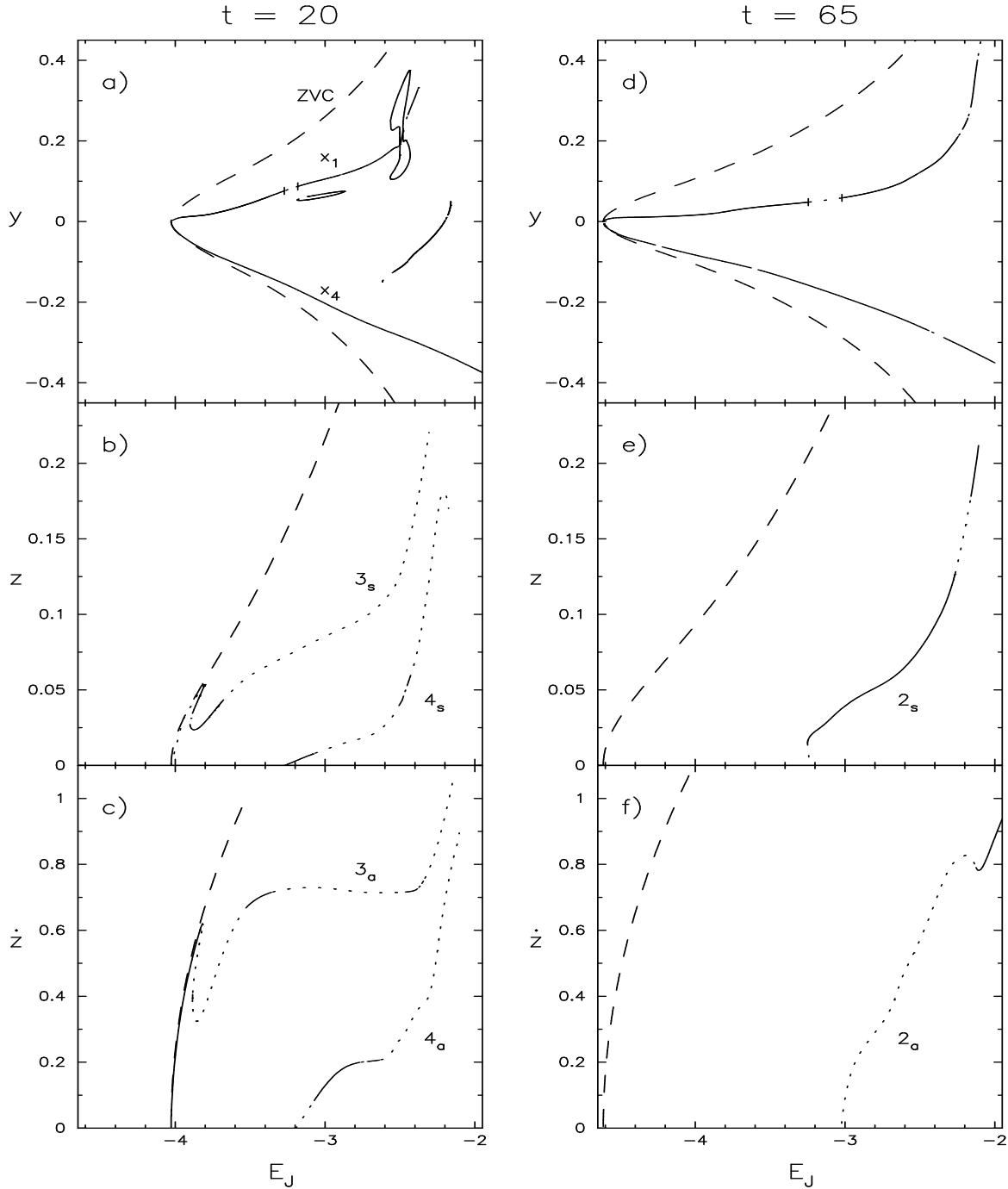}{18.2cm}{0}{90}{70}{-275}{-10}
\caption{Characteristic diagrams for purely collisionless model~A.
Stable sections are represented by solid lines while unstable are
broken. The dashed curve is the zero velocity curve. The families
consist of symmetric and anti-symmetric \OrbFam{2}{n}{1} orbits,
where the value of $n$ is given in the diagram for each characteristic.}  
\end{figure*}
\begin{figure*}
\plotfiddle{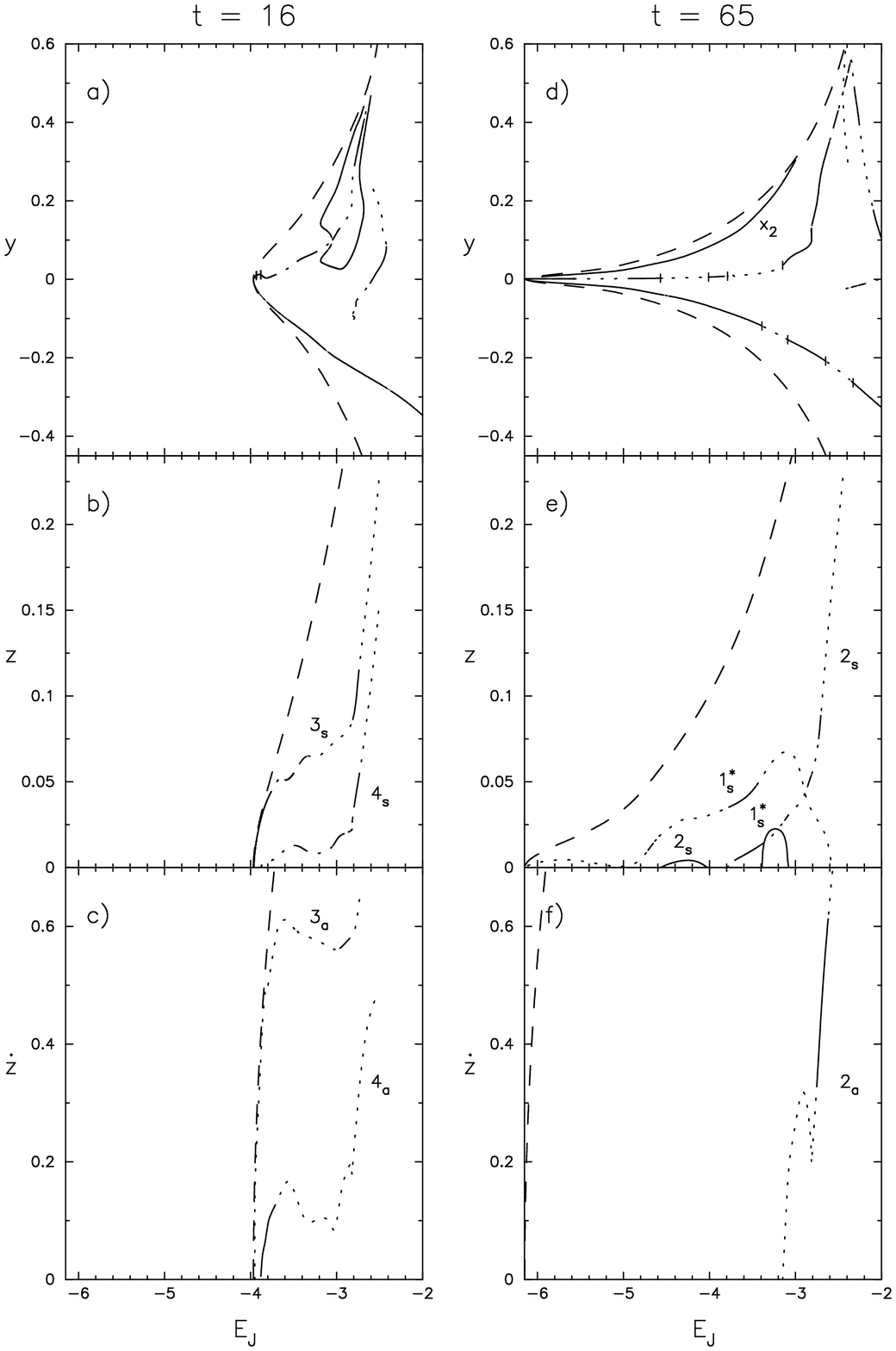}{18.2cm}{0}{90}{70}{-275}{-10}
\caption{Characteristic diagrams for stellar+gas  model~B.
Stable sections are represented by solid lines while unstable are
broken. The dashed curve is the zero velocity curve. The families
consist of symmetric and anti-symmetric \OrbFam{2}{n}{1} orbits,
where the value of $n$ is given in the diagram for each characteristic.
The asterisk indicates a retrograde family.}
\end{figure*}
The dashed curve denoted by the symbol ZVC is the
zero velocity curve, which delineates the accessible region in the
plane based on energy considerations.  Stable sections of the
characteristics are represented with a solid line, while unstable
sections are broken.  
The two main families have been labeled according
to the notation of Contopoulos \& Papayannopoulos \shortcite{con80}.
The corotation radius is at $\Ej=-2.18$ or approximately 9.2\,kpc, as
defined by the location of the Lagrange point in the orbit model potential.

The family labeled $x_1$ are the orbits that predominately give the bar
its structure.  They are elongated along the bar and orbit in the same
sense as the bar (direct).  In the rotating reference frame of the bar
they move in and out, twice for once around the center (2/1).
At this time, as one follows the characteristic from the center out
to higher energies, the orbits become more and more pointy, until
at around $\Ej=-3.62$ they develop small loops along the bar axis.
As the orbits grow in size, so do the loops, until a little 
before the turn-up in the characteristic, they shrink
and disappear.  At this point, the orbits extend to about 8\,kpc
along the bar and have an axis ratio of 1:4.  Past the bend at $\Ej=-2.5$,
the orbits continue to grow in size and become ever
increasingly more rounded.  Some examples of these orbits are shown
in Figure~3.

Bifurcating from the $x_1$ sequence near the bend
is a bubble consisting of orbits which transition from 2/1 to 3/1 to 4/1 as
one moves along it to higher energies.  A similar bubble was found by
Pfenniger \shortcite{pfe84}, except here most of the orbits have loops
and the extreme 4/1 orbits have a more pronounced diamond shaped.  There
is also a detached loop of 2/1 direct orbits under the $x_1$ sequence
from $\Ej=-3.19$ to $-2.86$, which covers a range of semi-major axes
from 4.2 to 5.5\,kpc.  The orbits on the upper part of this loop are
quite pointy, while those along the bottom have more rounded ends.
The two sections at either end of the upper segment are horizontally
unstable.

Also in Figure~4a is present the family labeled $x_4$.  These are 2/1
orbits as well, but move in a retrograde direction and are slightly elongated
perpendicular to the bar.  The characteristic is stable along 
most of the length shown in the figure, except for a few regions
at lower energies where it is horizontally unstable.  
This family can play only a minor role at this early time, since a 
significant population of retrograde orbits would stabilize the disk
against the formation of a bar (e.g., Kalnajs 1977; Zhang \& Hohl 1978;
Christodoulou, Shlosman \& Tohline 1995).

Bifurcating from a vertically unstable section of the $x_1$ 
characteristic are two 3D orbit families.  One
family bifurcates in $z$ from the low energy edge of the strip at
$\Ej=-3.26$, the other in $\dot{z}$ from the high energy side at $\Ej=-3.20$. 
These orbits make two radial oscillations in the $(x,y)$ plane
and four vertical oscillations in $z$ as the orbit closes one rotation
about the center. The two families form a pair, with
the bifurcations in $z$ and $\dot{z}$ being, respectively, symmetric and
anti-symmetric about the $(y,z)$ plane.  
Using the notation of Sellwood \& Wilkinson \shortcite{sel93},
these orbits are given as \OrbFam{2}{4}{1}, with an `s' or `a' subscript
added to indicate the sense of symmetry.  These orbits have the same
shape, in projection onto the $(x,y)$ plane, as the $x_1$ orbit of the
same energy, including the loops.
The symmetric family is stable near the bifurcation point, where they have
semi-major axes of 4\,kpc, but become unstable over much of the rest
of the characteristic.  The anti-symmetric family is also unstable
over most of its characteristic, except for a section between
$\Ej=-3.07$ to $-2.64$.  The maximum $z$ extent of the stable orbits of 
these two families is less than 300\,pc.

There also exists a pair of \OrbFam{2}{3}{1} families apparently 
bifurcating from the origin. However, one should keep in mind that
the gravity has been softened at the center.  Both the symmetric
and anti-symmetric family stay very close to the zero velocity curve
until the orbits extend some 0.5\,kpc above the disk plane, then 
move away, following an abrupt turn-down in their characteristics.
The symmetric family is mostly unstable, except for a section following
the turn-down.  The anti-symmetric family is stable before the 
turn-down and mostly unstable following it.

It should be noted that no $x_2$ orbits or symmetric \OrbFam{2}{2}{1}
family corresponding to planar and vertical ILR respectively, are
present at this time.

\subsubsection{Periodic Orbits at $t=65$}

Figure~4d shows the same model at $t=65$.  At this later time, the 
potential has deepened due to the mass redistribution in the system
following bar and buckling instabilities. The corotation
radius has moved slightly out to $\Ej=-1.92$ or about 11.5\,kpc.
Because the bar has weakened, the planar characteristics are simpler
than before, with the absence of the bubbles and detached loops.
The $x_1$ orbits are stable from the center out to $\Ej=-3.24$, 
where a vertical instability strip begins.
At this point the orbits extend to about 3\,kpc along the bar, with
an axis ratio of 1:6.  At $\Ej=-3.02$ the orbits once more become
stable and remain so up to near where the characteristic turns upward
toward the zero velocity curve.  The orbits along this section do 
become slightly more pointy before starting to fatten into a more oval
shape, but never develop loops.  The steep section of the characteristic,
past the turnup, has several sections of varying stability.  Because
these bifurcation points are near the increasingly chaotic region 
of the corotation, we have made no attempt to further explore them.
No $x_2$ orbits have been detected, hence no radial ILR(s) exist.

The $x_4$ characteristic is also shown and is stable along the entire
length displayed, except for a period-doubling horizontal bifurcation
near $\Ej=-3.62$.

 From the vertically unstable strip on $x_1$, bifurcate two families of 
\OrbFam{2}{2}{1} orbits: a symmetric family (BAN) in $z$ at $\Ej=-3.24$ and
an anti-symmetric family (ABAN) in $\dot{z}$ at $\Ej=-3.03$.  The symmetric 
family is unstable near its bifurcation point, but becomes stable following
a bend in the characteristic, and remains so along most of the curve
shown in the diagram up to $z\sim2$\,kpc. In contrast the anti-symmetric
family is unstable over most of the characteristic shown, becoming stable
only for $\Ej>-2.10$. The symmetric family bifurcating from $x_1$ at the
instability strip are characteristic of a vertical ILR 
(Pfenniger 1984; Friedli \& Pfenniger 1991).  

\subsection{Model With Gas}

\subsubsection{Periodic Orbits at $t=16$}

While the bending instability is operating, the disk is
highly asymmetrical in the vertical direction.  The interpretation
of the 3D orbit families, computed from the symmetrized potential at
such a time would be highly ambiguous.
Since the onset of the bending instability occurs earlier for the
model with gas, slightly before the time of maximum bar strength,
we are required to perform the orbit analysis at an earlier time
than for model~A.  Nonetheless, in the following section we will
examine some of the more robust planar features of this model at
full bar strength ($t=20$) as well.

For the full analysis we chose the time $t=16$, at which the bar is at
about 80\% of its maximum strength and before the onset of the large gas
inflow rate.  The characteristic diagram for model~B at $t=16$ is
shown in Figure~5a.
At this time the corotation radius is at $\Ej=-2.48$ or about 7.2\,kpc.

The $x_1$ orbits between $\Ej=-3.87$ and $-3.63$, or semi-major axes between
1-2\,kpc are pinched perpendicular to the bar into a pickle-like shape.
An example of such an orbit is shown in the panel labeled B16 of Figure~3.
At this time, the orbits do not become pointy or develop loops. 
Similar to model~A at $t=20$, there is a bubble of 2-3-4/1 orbits
bifurcating from the $x_1$ sequence near the bend.  Here however, the
orbits cover a larger range of energy and there exists an additional
smaller bubble enclosed within the larger. No detached
loop of 2/1 orbits and no $x_2$ orbits were found.

Along the $x_1$ sequence there are several regions of instability. Three are
horizontally unstable sections from $\Ej=-3.72$ to $-3.43$, $\Ej=-3.25$ to
$-3.05$, and covering the bend from $\Ej=-2.97$ to $-2.80$.  There are also
two short vertically unstable sections near the center. Bifurcating
from the plane near $\Ej=-3.96$ are a pair of \OrbFam{2}{3}{1} orbit
families.  Similar to model~A, these families at first stay very close
to the zero velocity curve.  The symmetric family is stable near the 
bifurcation point, but
then alternates a few times between stable and unstable sections, while
the anti-symmetric family is almost everywhere unstable.  Also from
near the center, at $\Ej=-3.88$, bifurcate a pair of \OrbFam{2}{4}{1}
families.  The symmetric family is unstable over most of its characteristic,
including near the bifurcation point, and has a complex unstable section
following the sharp turn-up at $\Ej=-2.82$.  The anti-symmetric family
is stable from it bifurcation point to $\Ej=-3.72$, after which it
remains unstable.

\subsubsection{Periodic Orbits at $t=20$}

For comparison with the planar orbits of model~A, we have also performed 
an orbit analysis of model~B at $t=20$, the time at which the bar
has reached maximum strength.  The characteristic diagram is shown
in Figure~6.
\begin{figure}
\plotfiddle{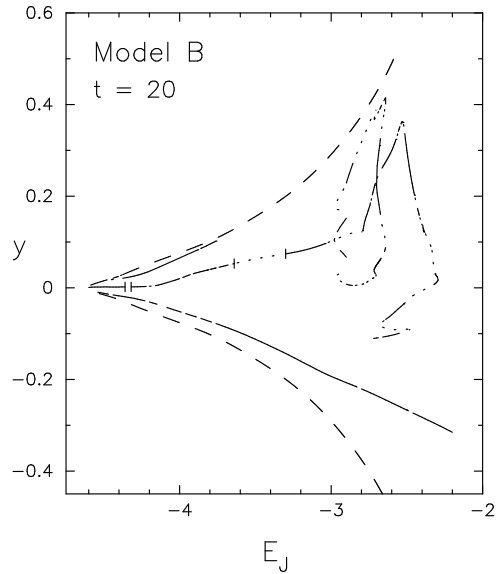}{7.3cm}{0}{60}{60}{-120}{-210}
\caption{Characteristic diagrams for stellar+gas model~B at $t=20$.
Stable sections are represented by solid lines while unstable are
broken. The dashed curve is the zero velocity curve. At this time
the bar is near maximum strength and an ILR has appeared.}
\end{figure}

At this time the corotation radius is at $\Ej=-2.38$ or about 8.2\,kpc,
somewhat smaller than the purely collisionless model~A.
Similar to model~A, loops are present in the $x_1$ orbits at this time,
as is a bubble of 3/1 orbits.  However, in contrast to model~A,
a family of $x_2$ orbits already exists. The gas inflow
and the deepening of the central potential are responsible for their
appearance.

The $x_2$ family, like $x_1$, are also direct 2/1 orbits, but elongated
perpendicular to the bar.  Their presence is indicative of an ILR(s) in the
nonlinear regime.  The characteristic covers a range in energy
from $\Ej=-4.45$ to $-3.83$, along which the orbits extend perpendicular
to the bar from 0.18\,kpc to 0.92\,kpc, with axis ratios of 1:2 and
1:3, respectively.  

\subsubsection{Periodic Orbits at $t=65$}

By time $t=65$ the potential well of model~B has greatly deepened, a
reflection of the steady accumulation of gas in the nuclear disk.
The corotation radius has moved outward, though not as much as in
model~A, to $\Ej=-2.25$ or 9.2\,kpc.  
Unlike model~A, the $x_1$ family, once having obtained loops, has
retained them throughout the remaining evolution.  Compared to the earlier
time the loops are now longer, though less fat. The sections of
instability have grown considerably during the evolution and at this
time the $x_1$ orbits are mostly unstable, except in the inner regions.

The 4/1 bifurcation from the $x_1$ family (UHR) occurs at $\Ej=-2.36$, at
a point where the $x_1$ orbits have an extent along the bar of 
around $R_{\rm UHR}=6.4$\,kpc and an eccentricity of $e=0.5$. 
This corresponds
to the approximate location of the outer edge of the gas ring
described in section~4.2.  Most of the 4/1 orbits are either unstable or
have loops and the gas appears to have accumulated on $x_1$ orbits with
$\Ej$'s just below the 4/1 gap.  If we take $R_{\rm UHR}$ to be indicative
of the maximum possible bar length, then the value of 
$R_{\rm CR}/R_{\rm UHR}=1.4$.  This value is within the limits of the
empirical range given by Athanassoula \shortcite{ath92}.

The extent of the $x_2$ family has greatly increased from $t=20$
and now covers a range of energy from essentially the minimum
to $\Ej=-3.0$, with an axis ratio of around 1:3. 
The family is stable along its entire length.

The $x_1$ family also has two sections of vertical instability before
the sharp bend upwards at around $\Ej=-2.8$.  The first has it's two
ends at $\Ej=-4.52$ and $-4.03$ connected by a loop of stable symmetrical
\OrbFam{2}{2}{1} orbits (BAN), which bifurcate
from the plane in $z$. These orbits are stable everywhere,
span a range along the bar major axis of 0.06--1.1\,kpc, have a 
maximum vertical extent of about 100\,pc, and have loops in projection
onto the $(x,y)$ plane.

 From the second strip also bifurcate a pair of \OrbFam{2}{2}{1} families. 
The symmetric family, which bifurcates in $z$ at $\Ej=-3.78$, is unstable 
between its bifurcation point and $\Ej=-3.07$, then alternates
three times between stable and complex unstable sections until
$\Ej=-2.70$, after which it remains predominantly unstable, unlike in model~A.
The anti-symmetric family, which bifurcates in $\dot{z}$ at $\Ej=-3.15$,
is unstable between its bifurcation point and $\Ej=-2.75$, after which
it becomes stable.

The $x_4$ family has two sections of vertical instability.  The
one at lower energy has its two ends at $\Ej=-3.39$ and $-3.08$, connected
by a stable vertical family of symmetric \OrbFam{2}{1}{1} orbits that
bifurcate from the plane in $z$, and which 
are inclined to the y-axis (${\rm ANO}_y$).  These orbits have radii
between 1.1-1.5\,kpc.  From the higher energy strip also bifurcate a
pair of \OrbFam{2}{1}{1} families.  The symmetric family, which
bifurcates in $z$ at $\Ej=-2.58$, is everywhere unstable except for two
small sections.  Similar to the loop, these orbits are inclined 
to the y-axis.  This family has been followed to near the center, well
within the softened inner region, and is consistent with the family having
bifurcated from the origin.

\subsubsection{Surface of Sections at $t=65$}

Most orbits in a galactic-like potential are not periodic.
However, the periodic orbits play an important role, in that
they trap regions of phase space about them.  These trapped regions
are referred to as regular regions and the orbits which reside in them
are called regular orbits. The motions of these orbits
are confined to a two dimensional surface referred to as an invariant
tori and which surround the parent periodic orbit.  Orbits which are not
trapped are called irregular. Unlike the regular orbits, they are not 
restricted to a sub-surface and may wander throughout the non-regular 
regions of phase space, at least within energy considerations.

To determine the extent of phase space trapping about
the stable periodic orbits, we shall examine the surface of
sections (SOS).  These diagrams have been constructed
by integrating orbits of a given $\Ej$ in the {\it unsymmetrised}
potential, and marking a point in
the $(y,\dot{y})$ plane each time it crosses the line $x=0$ with
$\dot{x}<0$ (e.g., Binney \& Temaine 1987).  An example of such diagrams 
for a barred potential can be seen in the top left panels of Figures~7
(Model~A) and 8 (Model~B), for $\Ej=-4.4$ and $-5.4$, respectively.
\begin{figure*}
\plotfiddle{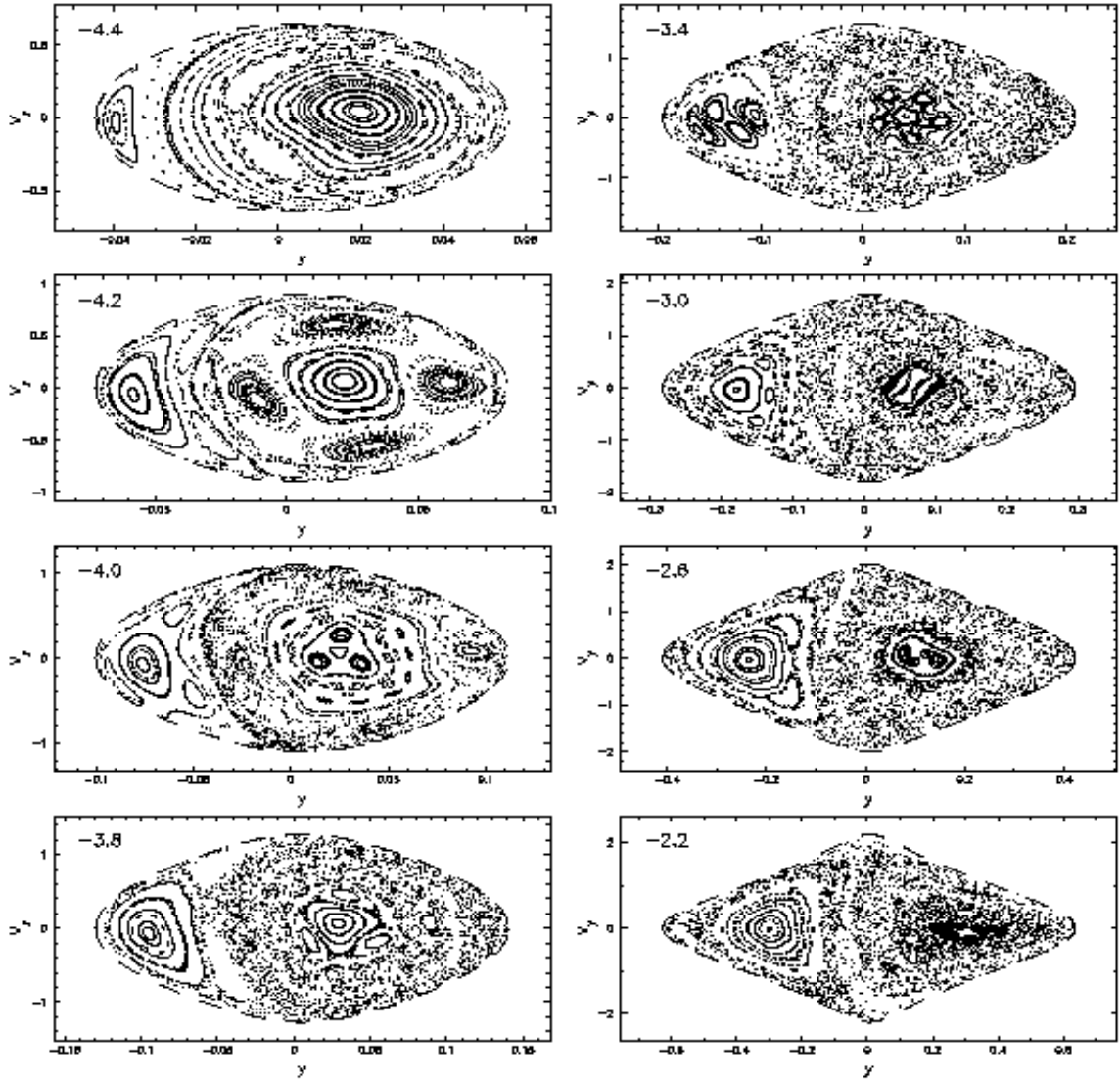}{17cm}{0}{90}{85}{-275}{-150}
\caption{Surface of section diagrams for model~A at $t=65$.
The direct $x_1$ and retrograde $x_4$ families dominate the phase space.
At higher energies, as the corotation radius is approached, the
fraction of phase space which is stochastic increases.}
\end{figure*}
\begin{figure*}
\plotfiddle{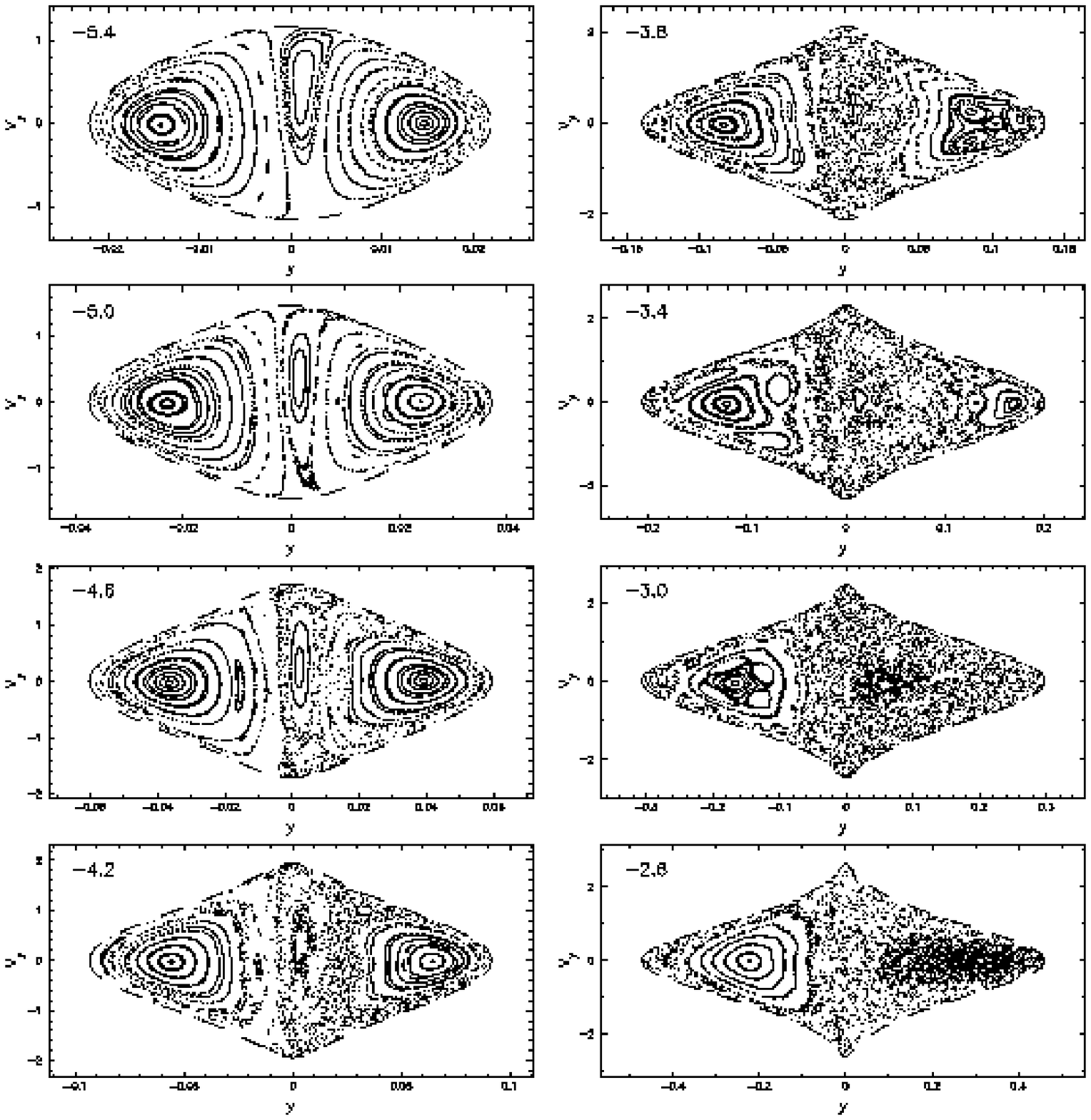}{17cm}{0}{90}{85}{-275}{-150}
\caption{Surface of section diagrams for model~B at $t=65$.
No significant support for the bar, in terms of regular orbits, 
exists at this time. The regular regions of phase space are 
dominated by the $x_2$ and $x_4$ families. At energies where the $x_1$
family traps a non-negligible fraction of the available phase space
(e.g. $\Ej=-5.4$ and $-5.0$), the orbits are significantly inclined
to the bar major-axis and provide little support.  At higher energies
the stochastic regions grow at the expense of the regular regions.}
\end{figure*}
The regular orbits in these diagrams form the closed curves which surround
the fixed points of the parent periodic orbits at the centers.  These
curves are referred to as invariant curves and are a cross section
of the invariant tori.  The left side of each diagram, with $y<0$, 
represents retrograde orbits, while the right side, with $y>0$,
represents the direct or prograde orbits.  For the above diagram frames
chosen as examples, there are two regular regions in Fig.~7, which from
left-to-right, are associated with the $x_4$ and $x_1$ orbital families.
In Fig.~8, there are three
regular regions, which from left-to-right, are associated with the 
$x_4$, $x_1$, and $x_2$ orbital families.

Figure~7 shows that the $y > 0$ regular regions of the phase space in the
model~A are dominated by the $x_1$ family which dissolves at around
$\Ej=-2.4$.  In contrast, Figure~8 shows that most of the regular regions of
phase space in model~B are dominated by the $x_2$ and $x_4$ families.
At lower energies, where the $x_1$ family occupies a non-negligible
fraction of the available phase space, the fixed point for the
family is at $\dot{y}>0$, indicating that these orbits are inclined
at an angle oblique to the bar major-axis in a trailing direction.
The general trend of the offset angle is of increasing value as
the center of the potential is approached.
As one moves out in energy, the fraction of phase space which is
regular decreases, with the $x_1$ family disappearing by $\Ej=-4.2$.
At higher energies, as the corotation radius is approached, the
stochastic regions continue to expand ($\Ej=-3.8$ and $-3.4$), 
with the $x_2$ family completely dissolving by about $\Ej=3.0$. 
The regular retrograde orbits continue to fill up most of the available
phase space for $y<0$, so stochastic orbits are not as important here.
No significant support for the bar, in terms of regular orbits, appears to
exist at this time.  

A picture library of two and three dimensional periodic orbits that
were found in these models is given in Figures~3 and 9.

\begin{figure}
\plotfiddle{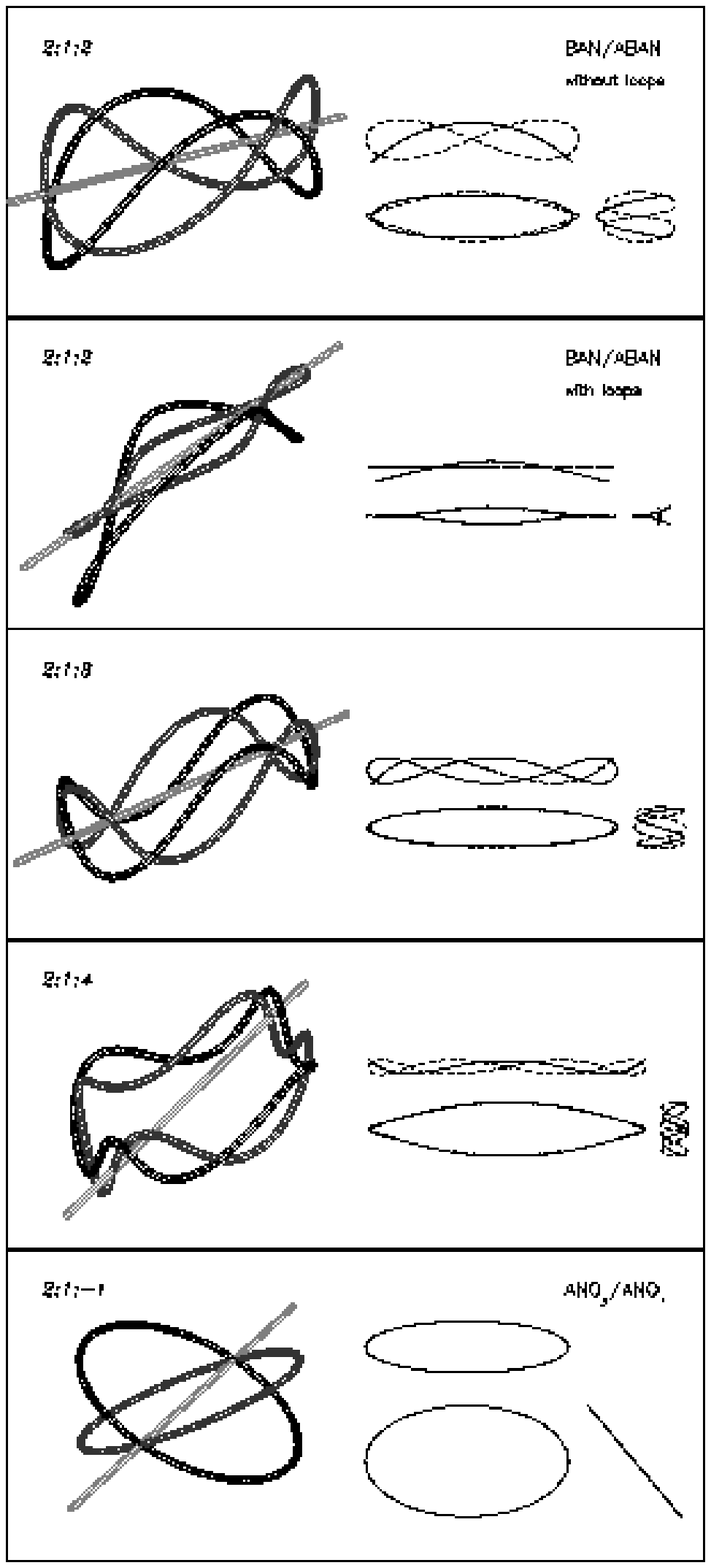}{19.5cm}{0}{75}{80}{-242}{-80}
\caption{Some examples of 3D orbits from models A and B.  Shown
are the symmetric and anti-symmetric orbital pairs of the same $\Ej$
and their projections onto the fundamental orthogonal planes. The
x-axis is indicated in the 3D figures by the cylindrical bar.}
\end{figure}

\section{DISCUSSION}

The diverging evolution of both models, which started with 
nearly identical initial conditions, can be attributed solely
to the gravitational effects of gas redistribution in the galactic
plane. We analyze two snapshots of the model potentials, the first 
one at an early time corresponding to the maximal strength of the 
non-axisymmetric bar potential, $t=20$, and the second one at a time when 
dynamical evolution of the models has slowed down substantially, $t=65$. 
Although the models differ already at $t=20$, because a nonnegligible
fraction of the gas already found its way to the center, these
differences become much more pronounced at the advanced stages of
evolution.

\subsection{Morphological Differences}

After ten rotations, the morphological differences in the bar region of
both models are clearly visible. Even earlier, such dynamical
parameters as the bar strength (defined by its axis ratios), rotation curves 
and density profiles show growing differences delineating dissimilar
underlying dynamics. This is evident in the development of the 
two major instabilities in these models, the bar instability and the subsequent 
bending instability. The general trend is that the presence of the 
self-gravitating gas component in the disk makes both instabilities 
substantially milder, in line with our previous results.

In general, the evolution is accelerated
in the B model compared to A, even before the bar reaches its maximal strength
(which happens simultaneously in both models). The bar pattern speed in B 
is substantially higher than in A after approximately one rotation and 
this difference increases with time. 
We note also that the pattern speeds of
the stellar bars show different behavior after the initial quick decline ---
due to the continuous gas inflow to the center, the bar in model~B slowly
increases its speed, in contrast to that of the bar in model~A which
continues its secular slowdown because of the interaction with the halo
and the outer disk.
An additional signature of the accelerated evolution of model~B is the early
onset of the bending instability and the sudden increase in the vertical
scaleheight in the inner few kpc of the disk.

The central potential of model~B substantially deepens due
to the gas inflow, resulting in the formation of the $x_2$ family of orbits
which delineate the radial ILR. Fig.~5 shows 
that the inner resonance region is very
broad and extends almost from the center (i.e., from IILR), to 3\,kpc in radius
(OILR), which encompasses the minor axis of the bar. The inner $x_2$ orbits 
in the vicinity
of the IILR are heavily populated with gas. The stellar bar weakens
as the anti-aligned $x_2$ orbits become more important. At the peak of
its strength the $m=2$ mode amplitude is larger by $\sim50$\% in model~A, and 
the difference grows by a factor of two by $t=65$.
Towards the end of the simulation, at $t=100$, the B model bar almost ceases 
to exist, having weakened by a factor of ten.

We note, that the 
gravitational potential is sufficiently softened at the position of the
IILR and so the dynamical effects of this resonance are eliminated. This
result contrasts the modeling of inner disk in M100 by Knapen et al. 
\shortcite{kna95}. 
In the latter work, the spacing between the ILRs is much smaller and the
position of the IILR is further away from the center, leading to the formation
of a nuclear ring just outside the IILR. Both numerical simulations, the
one presented here and by Knapen et al., underline two main alternatives
in the evolution of gas distribution in the circumnuclear regions of
disk galaxies. Further work is required to understand the effects
of the IILR on the galactic morphology.

Another important morphological difference between models A and B can
be found in the properties of the central bulge which forms as a result
of vertical buckling in the stellar bar. In particular, the gas model
gives a bulge-to-disk ratio and bulge shape value, as defined by eq.(1),
that is more consistent with the observed distribution.  Moreover, as the
gas falls toward the center, the change in value of the bulge shape
parameter, is consistent with an evolution toward an earlier morphological
type.

\subsection{Dynamical Differences}

The morphological differences between the models is a result of more
fundamental differences in the orbital dynamics, in particular, from a 
change in the relative importance of various orbital families in the
overall phase space available to the system. First, as the SOSs show
(Fig.~8), model~B at $t=65$ has a 
larger fraction of the phase space populated by chaotic orbits. This 
comes at the expense of the $x_1$ orbits, which are always the most important 
orbits supporting the bar. The $x_1$ family itself becomes largely 
unstable (Fig.~5d), and although remains
present everywhere inside the CR, it occupies only a small fraction of the 
phase space and is unable to trap substantial numbers of regular orbits.
Hence, at these late times the $x_1$ family is dynamically unimportant.
Such a behavior was also observed by Hasan, Pfenniger \& Norman 
\shortcite{has93} using an analytical potential with various degrees
of central concentration. 

Surface of sections of the unsymmetrized potential of model B have
shown the presence of the innermost oblique
$x_1$ orbits during the later times of its evolution and which dominate
the phase space at the lowest energies (Fig.~8). These orbits trail
the major axis of the stellar bar and are mostly perpendicular to the 
major axis of the gaseous nuclear disk which leads the stellar bar
by about $80^\circ$.  They do not contribute to the support of the bar. 
The origin of this anomalous oblique family of $x_1$ orbits 
is explained by the fact that at these low energies (and small distances
from the
center), the non-axisymmetric gravitational potential of the large-scale bar
is barely felt, and the Laplace plane of the potential is instead controlled
by the nuclear disk. 

We also observe that most of the $x_1$ orbits in model~B posses loops 
or appear ``pointed'' within the 5\,kpc region of the stellar bar, while
$x_1$ orbits in model~A are round everywhere. This implies grave
consequences for the gas dynamics.  Because intersecting orbits can not
support a steady state gas flow, the gas loses energy and angular momentum, 
and moves inwards (as indeed is observed in this model). 

By changing the mass distribution in the galactic plane, the gas also
affects the vertical structure of the disk by creating vertically unstable
gaps in the planar (2D) orbits as well as by destabilizing the 3D 
families bifurcating from the plane, as discussed in section~5.
The symmetric (BAN) and antisymmetric (ABAN) \OrbFam{2}{2}{1}
families of orbits have the largest extensions away from the disk plane
and appear to dominate the phase space (Figs.~4 and 5).
As Pfenniger \& Friedli \shortcite{pfe91} found, these families
of 3D orbits bifurcate at the vertical resonance gap in the plane. 
The vertical ILR is typically found in the neighborhood of the radial
ILR (Hasan et al. 1993). 
In our case this is true for model~B, which has the outer ILR at a
distance of about 3\,kpc from the center at $t=65$. Model~A has a vertical 
ILR in approximately the same place, but lacks the planar ILRs. 
No vertical ILRs are present during the early stages of the evolution. 

\subsubsection{3D Orbits}

The orbit analysis reveals that there exists a substantial difference between
the models in the stability of BAN orbits and in their extension above the
plane and into the halo. Both BAN and ABAN appear with the bending instability
and both are present in the models at $t=65$ (Figs.~4/5e,f). The bifurcation
points of these families moved inward during the evolution of model~B,
especially that of the BAN orbits. The BAN in model~A, however, are stable
up to approximately 2\,kpc above the disk, while in the presence of the
gas (model~B) they are destabilized above $z\sim500$\,pc. The ABAN are
unstable to large distances above the disk plane. The fact that in  
model~B the stability region of these families shrinks and is limited to
a narrow layer above the disk plane, is important if these families are
related directly to the bending instability and the formation of the 
peanut shape bar profile (Pfenniger \& Friedli 1991;
Sellwood \& Wilkinson 1993).  This means that the vertical
bending in the bar will be (at least partially) damped in the presence
of the gas and explains why the bending is substantially milder and reaches
a smaller amplitude in the B model as compared to A (Figs.~1 and 2). 
The breaking of the z-symmetry of the bar, clearly observed in Fig.~1 at
$t=39-40$, temporarily eliminates the planar and populates the BAN orbits,
giving the bar its characteristic peanut shape in model~A. This effect is 
partially suppressed in the B model, both because the z-symmetry is 
maintained to a higher degree, as seen in Fig.~2, and because the BAN
orbits are largely unstable (except in the immediate vicinity of the disk
plane), and so the particles which leave the planar orbits and enter into
the halo are populating chaotic orbits. The peanut shape is barely observed
in this case.

\section{CONCLUSIONS}

We have compared the diverging evolution of a two-component (gas$+$stars) 
galactic disk
embedded in a live halo with that of an nearly identical pure stellar disk.
The models have been chosen to be bar unstable and developed
a large-scale stellar bar in a few dynamical times, which in turn experienced
vertical buckling. The resulting
non-axisymmetric potential has led to the loss of angular momentum by
the gas and its redistribution towards the center, along with a
modification of the gravitational potential there. In the early stages of
the evolution no linear or nonlinear resonances are present in the inner
disk.

We confirm that the radial gas inflow leads to the formation of the OILR
resonance inside the bar. The IILR appears very close to the center and
its dynamical effect is limited by the gravitational softening. 
The single vertical
resonance, however, forms in both models with the growth of the bar mode.
The existence of these resonances was inferred from the orbits analysis and is
related directly to the appearance of particular 2D and 3D families of orbits. 
The bar is weakened by a lack of orbits supporting it.
In the plane, the region inside the OILR is mostly dominated by the
$x_2$ orbits which are elongated perpendicular to the bar, while the
$x_1$ orbits which are only found in the central regions are at oblique
angles to the bar.  This model
does not lead to the formation of gaseous nuclear ring and provides an
interesting alternative to the evolution of gas distribution in the
circumnuclear regions of disk galaxies when the IILR is too close to the
center.

We find that the bar mode amplitude is substantially lower in the presence
of the gas at all times, and the amplitude decays steadily by a factor of 
2 on a 
time scale of $\sim 2\times 10^9$ yrs, or $\sim 7$ rotations. We also 
find that the buckling instability which
helps to populate the 3D \OrbFam{2}{2}{1} orbits due to the temporary
breaking of the z-symmetry in the disk, has a smaller amplitude in the
presence of the gas. In particular, the characteristic peanut-shape of the
inner bar is greatly weakened and `washed-out' due to destabilization of 
the 3D \OrbFam{2}{2}{1} orbits
in the more centrally concentrated model containing gas. The orbit
analysis shows that the increased
stability of the galactic disk should be attributed to the
larger population of chaotic orbits following the growth of the central
mass concentration to about 2\% of the galactic mass within 10 kpc. 
Milder buckling instability of the stellar bar has led to a much smaller
galactic bulge in better agreement with observations.

Our modeling supports the conjecture that the growth of central 
concentration in galaxies dissolves the main family of regular orbits
in the stellar bar and assists in the formation of a galactic bulge.
Overall, it hints about evolution towards more axisymmetric disks
and earlier morphological types. Taken at face value
one could expect a lower frequency of stellar bars in SOs, compared
to Sc's. Such a correlation is, however, not supported by optical and
near-infrared data. A resolution of this paradox is a matter for future work.

The characteristic time scale for bar dissolution appears to be short
in comparison with the Hubble time. Within this framework,
it is difficult to understand the large frequency, $\sim2/3$, 
of barred galaxies in the Universe. It is far from being clear
that galaxy interactions which (not always!) induce short-lived transient
bars, resolve this problem.

We conclude that the gas can dramatically alter the underlying
stellar dynamics of disk galaxies subjected to dynamical instabilities. 
The main consequences of radial gas redistribution in the disk is the 
increase of central mass concentration, and modification and destabilization 
of the main 2D and 3D periodic orbits. This results in a weaker stellar
bar and milder vertical buckling instability. The characteristic
peanut shape of the central regions in the bar is practically removed.
Overall this hints about a diverging evolution between initially gas-rich and
gas-poor barred galaxies.

\section{ACKNOWLEDGMENTS}

We are grateful to Rainer Spurzem and Christian Theis for providing both
time and support on the GRAPE hardware, and the referee, L. Athanassoula, 
for her valuable comments. The GRAPE 3Af special purpose
computer in Kiel was financed under DFG grant Sp 345/5-1, 5-2. 
C.H. acknowledges support from DFG grant Fr 325/39-1, 39-2. 
I.S. acknowledges support from NASA grants WKU-522762-98-6 and NAG5-3841.

\end{document}